\documentclass[runningheads]{llncs}
\usepackage{graphics,graphicx,subfigure,psfrag,mathtools,lineno,hyperref,amsmath,amssymb,dcolumn,bm,gensymb,color}
\begin{document}
\title{Methods of density estimation for pedestrians moving in small groups without a spatial
boundary}
%
\titlerunning{Estimating density without a spatial boundary}
%
\author{Pratik Mullick\inst{1,2}\orcidID{0000-0003-0133-0354}
\and
Cécile Appert-Rolland\inst{3}\orcidID{0000-0002-0985-6230}
\and
William H. Warren\inst{4}\orcidID{0000-0003-4843-2315}
\and
Julien Pettré\inst{1}\orcidID{0000-0003-1812-1436}
}
\authorrunning{P. Mullick et al.}
%
\institute{Univ Rennes, INRIA, CNRS, IRISA, Rennes, France
\and
Department of Operations Research and Business Intelligence, Wrocław University of Science and Technology, Wrocław, Poland
\and
Université Paris-Saclay, CNRS/IN2P3, IJCLab, Orsay, France
\and
Department of Cognitive, Linguistic and Psychological Sciences, Brown University,
Providence, Rhode Island, USA\\
}
\maketitle              
\begin{abstract}
For a group of pedestrians without any spatial boundaries, the methods of density estimation is a wide area of research. Besides, there is a specific difficulty when the density along one given pedestrian trajectory is needed in order to plot an `individual-based' fundamental diagram.
We illustrate why several methods become ill-defined in this case. We then turn to the widely used Voronoi-cell based density estimate. We show that for a typical situation of crossing flows of pedestrians,
Voronoi method has to be adapted to the small sample size.  We conclude with general remarks about the meaning of density measurements in such context.

\keywords{Crowd motion  \and Density estimation \and fundamental diagram}
\end{abstract}

\section{Introduction}

From the point of view of crowd management, understanding the fundamental relation between crowd density and speed, i.e. fundamental diagram (FD), is extremely useful.  Also for simulation based findings, FD acts as an essential tool to evaluate the capacity of simulations to predict realistic pedestrian flow. Thus, to construct a realistic FD, an effective method of density estimation is required.

In the context of self-organizing behaviour of human crowd motion, methods of density estimation are an important topic of research. Existing literature consists of a number of methods, however depending on the crowd situation a debate for the `best' method has not yet been resolved. Most of the research have been focused on situations where the moving crowd is constrained within a physical boundary. For such cases, density estimation using Voronoi tesselation \cite{DUIVES2015162,seyfried2010,cecile_vor,zhang2011b,cao2017} and a grid-based measure called XT method \cite{EDIE1963,DUIVES2015162,bode_c_h2019} have been reported to work well. However, there lacks a well defined method of density estimation for groups in an unbounded space.

A further difficulty arises when the fundamental diagram, instead of being plotted at a global or meso scale~\cite{seyfried2005}, is related to individual quantities~\cite{jelic2012a}. In one dimension, a proxy for the density can be the inverse of the spacial headway with the predecessor. In 2 dimensions, it is more difficult to find a good estimate of the density along the trajectory of a pedestrian.

For our research, we consider crossing flows of pedestrians without any spatial constraints, where two groups of people have to go across a sporthall from predefined initial positions such that they cross each other at specified values of the crossing angle $\alpha$. In the context of crossing flows, most of the previous investigations have been concentrated on $\alpha=90\degree$ \& $180\degree$. We performed experimental trials for a wide range of values of $\alpha$ with the goal to study crossing angle dependent properties~\cite{ploscb_pratik}. Currently, we want to investigate how the density-speed relationship for crossing flows is affected by the value of the crossing angle.

As an intermediate result, we remind here why several methods appropriate to determine a density field become inappropriate when it is the density along a pedestrian trajectory which is needed.
We discuss also how Voronoi method must be adapted for small groups without boundaries.
We shall end with a general discussion on the meaning of defining a density at such scales.

\section{Experimental details}

\begin{figure}[h!]
\includegraphics[width=\textwidth]{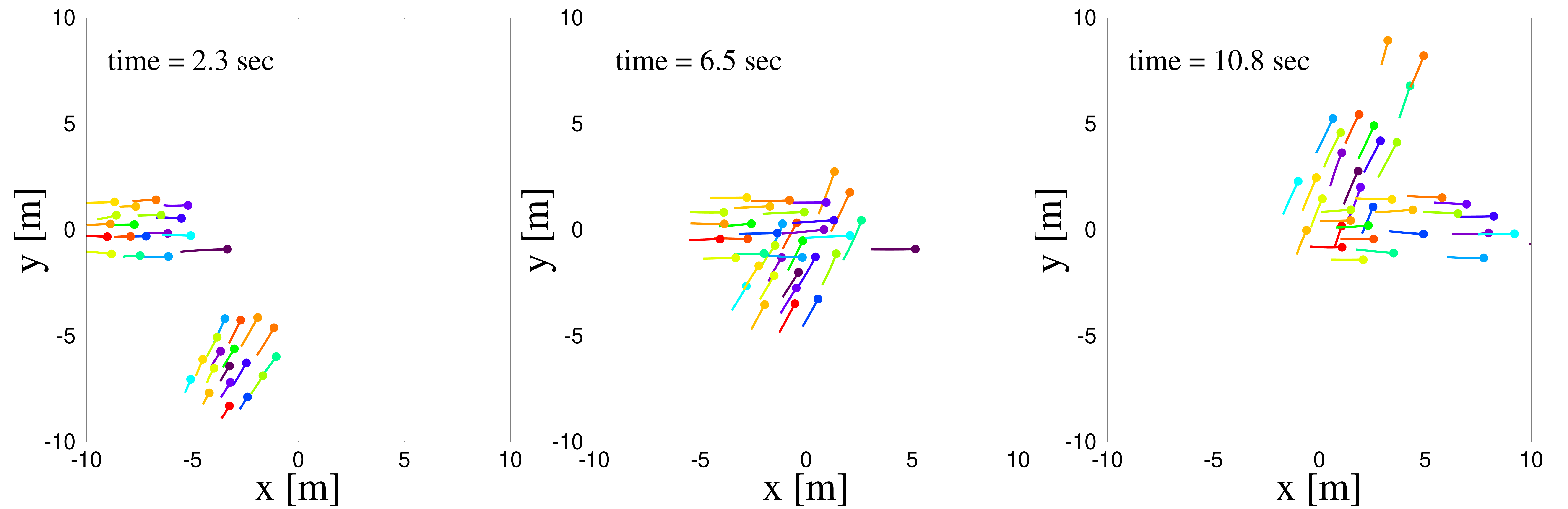}
\caption{Traces of all the pedestrians involved in a typical trial whose crossing angle is expected to be $60\degree$.
The dots represent the pedestrians and the tails behind each dot are the distances travelled by the pedestrian in previous 1.25 sec.} 
\label{traces}
\end{figure}

The data for crossing flows of pedestrians used for this research were obtained by experiments \cite{pedinteract_cecile,ploscb_pratik} using live participants on the campus of University of Rennes, France. Two different sets of volunteered participants (36 on Day 1, 38 on Day 2) were roughly divided into two groups (18 or 19 per group) and were instructed to reach the other side of a sport hall. Initial positions were prescribed so that the groups had to cross each other with seven different crossing angles ($0\degree$ to $180\degree$, at $30\degree$ intervals).
During each trial
we recorded the head trajectory of each pedestrian as a time series at a frequency of 120 Hz using a motion capture system based on infrared cameras (VICON).
In Fig. \ref{traces} we show the traces of all the pedestrians for a typical trial using filtered trajectories.

In a first analysis~\cite{ploscb_pratik}, we proposed two approaches to detect the stripes formed spontaneously at the crossing. The first one uses the whole dynamics of the crossing and determines individual stripes through an algorithm tracking how individual edges between pedestrian pairs can be cut - or not - during the crossing. The second requires a single snapshot at a given time, that is matched with a regular stripe pattern.


The analysis of the experimental data based on these two stripe detection methods has allowed us to show, first, that the average number of stripes decreases when the crossing angle increases.
Besides, we found a {\em squeezing effect}, meaning that during the crossing, stripes become thinner and thinner until their width reaches a minimum and expends again when pedestrian are close to the exit of the crossing region.

For further analysis, we would like to compute the fundamental diagram as measured during the crossing manoeuvre.
However, the fact that groups move in an unbounded space and are relatively small raises specific difficulties as far as density estimation is concerned.

\section{Estimating pedestrian density $\rho$ along a pedestrian trajectory}

A large number of methods have been developed to measure the density field of pedestrian flows. In order to determine the fundamental diagram at an individual scale, one has to associate a density to a specific pedestrian location. We shall see in this section that, while some methods do it without any difficulty, others completely fail once they are used along a trajectory.

We illustrate how this failure occurs using a classical grid-based method. The classical grid-based method to estimate the density of pedestrians follows an Eulerian approach. We divide the entire tracking region into square-shaped grids. For $d_g$ to be the length of these grids, the classical density $\rho_{\text{g}}$ is given by
    $\rho_{\text{g}}=n/{d_g}^2$,
where $n$ is the number of pedestrians that are located inside the grid. Note that sometimes, a single cell is used to measure density in a region of interest, for example in the crossing area of crossing flows~\cite{guo2010,zanlungo2023a,zanlungo2023b}.

\begin{figure}[h!]
\centering
\includegraphics[width=\textwidth]{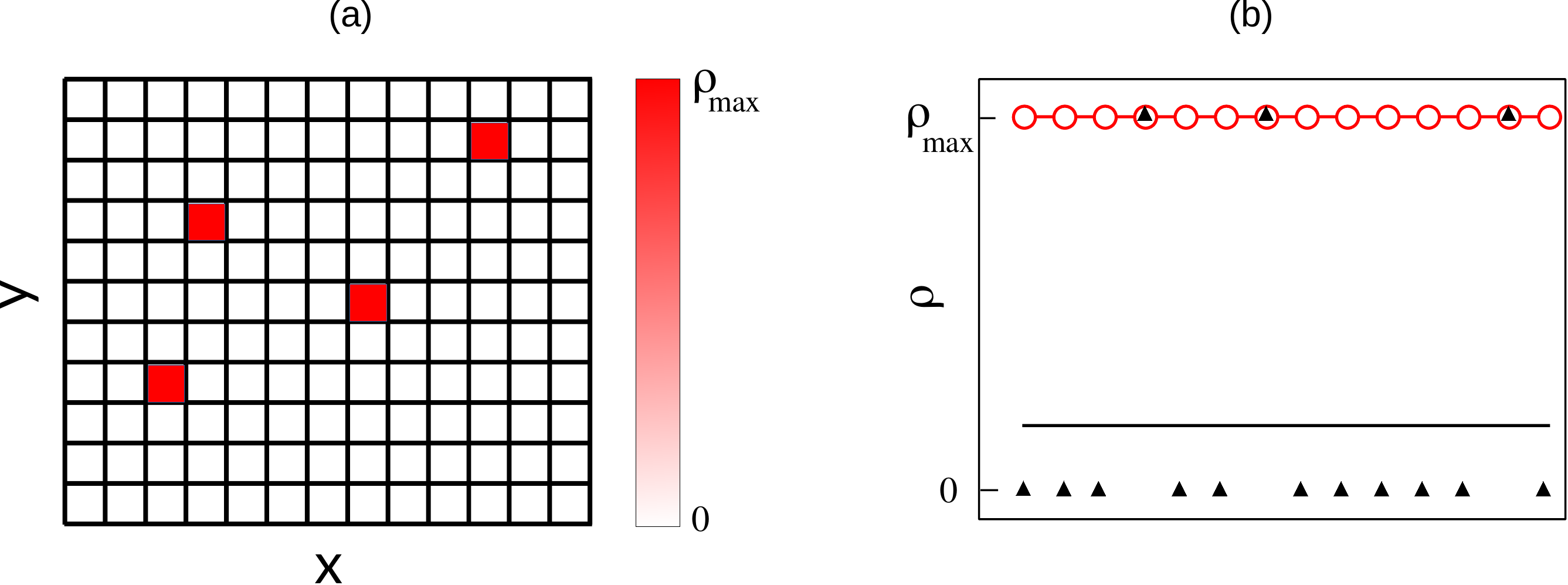}
\caption{Sketch representing the density variations in a grid-based method (a) when the whole density field is plotted, (b) for a time evolution in a fixed cell (black triangles) or along a pedestrian trajectory (red circles). Solid lines indicate the average values.}
\label{sketch_dens}
\end{figure}

As we are interested in small systems of pedestrians, $d_g$ must be small enough to capture the immediate vicinity of a given pedestrian.
In the limit of small cells, at most one pedestrian can be contained in the cell. The corresponding density is $\rho_{\text{max}}\equiv 1/{d_g}^2$.
If we plot the whole density field, we get typically a figure as in Fig. \ref{sketch_dens}(a) with some empty cells and some cells with density $\rho_{\text{max}}$.
The density field is not smooth, but using a spatial average, we get a reasonable density estimate - provided the area on which we average is homogeneous.
For stationary flows, it is also possible to average in time even for a rather small measurement area~\cite{zhang2011b} and to get average density values with rather low fluctuations.

If instead, we decide to store the density along the trajectory of a given pedestrian, the density in the cell where this pedestrian is located will always be $\rho_{\text{max}}$ - a value which can be arbitrarily high.
Any kind of averaging will always give $\rho_{\text{max}}$ (red line in Fig. \ref{sketch_dens}(b)), very far from the real density average value found previously (black line in Fig. \ref{sketch_dens}(b)). Actually 
density values are completely biased by the fact that we have selected cells conditionally, along the pedestrian trajectory.

Another variant could be to measure the density along the trajectory of a pedestrian without counting the pedestrian himself. But then again there is a bias: for small cells, the density along the trajectory would always be zero, far from the real density value.

This grid-based method could be used to estimate the local density around the pedestrian trajectory only if the cell size could be chosen large compared to the pedestrian interdistance, but small compared to density gradients. This is obviously not possible for the small pedestrian groups involved in our experiment.

A similar failure for density measurements along trajectories can be observed for other methods. The systematic bias and the fluctuations around this biased value have been studied in detail for 1 dimensional systems in~\cite{tordeux2015}. Here we illustrate it for two other methods in bidimensional space:

\begin{itemize}
    \item The XT method originally proposed by Edie \cite{EDIE1963} and modified by Duives et al \cite{DUIVES2015162}:
The density $\rho_{\text{xt}}(c,t)$ estimated in a cell $c$ of length $d_x$ and centered on pedestrian $p$ at time $t$ is given by
\begin{equation}
        \rho_{\text{xt}}(c,t)=\frac{\sum_q(T^q_{\text{end}}-T^q_{\text{begin}})}{{d_x}^2 \times T},
        \label{def_xt}
\end{equation}
where the summation is performed over all the pedestrians $q$ placed inside cell $c$ at time $t$. $T^q_{\text{begin}}$ and $T^q_{\text{end}}$ denote respectively the entrance and exit time in the cell of pedestrian $q$ with values bounded by $t-0.5T$ and  $t+0.5T$.
\item The Kernel method:
Each pedestrian contributes to the density field through a kernel $K_\textbf{h}$ such that
\begin{equation}
    \rho_\text{k}^\textbf{h}(\textbf{X}) = \sum^{N}_{i=1} K_\textbf{h}(\textbf{X}-\textbf{X}_i).
\end{equation}
An often used kernel is the Gaussian distribution function
$K_{h} = \frac{1}{2\pi h} e^{-\frac{1}{2 h^2}\textbf{X}^T \textbf{X}}.$
Along a pedestrian trajectory, the density value which is measured at the pedestrian location directly depends on $h$ through the normalization: $\rho_{max} = \frac{1}{2\pi h}$ and can thus be arbitrarily high for small $h$.
\end{itemize}

In these methods, the parameters $d_g$, $d_x$, or $h$ determine the typical spatial length on which the presence of a pedestrian has an effect on the density when measuring the density field. It also determines the density value at the pedestrian location when density is measured along a trajectory.

\begin{figure}[h!]
\centering
\includegraphics[width=\textwidth]{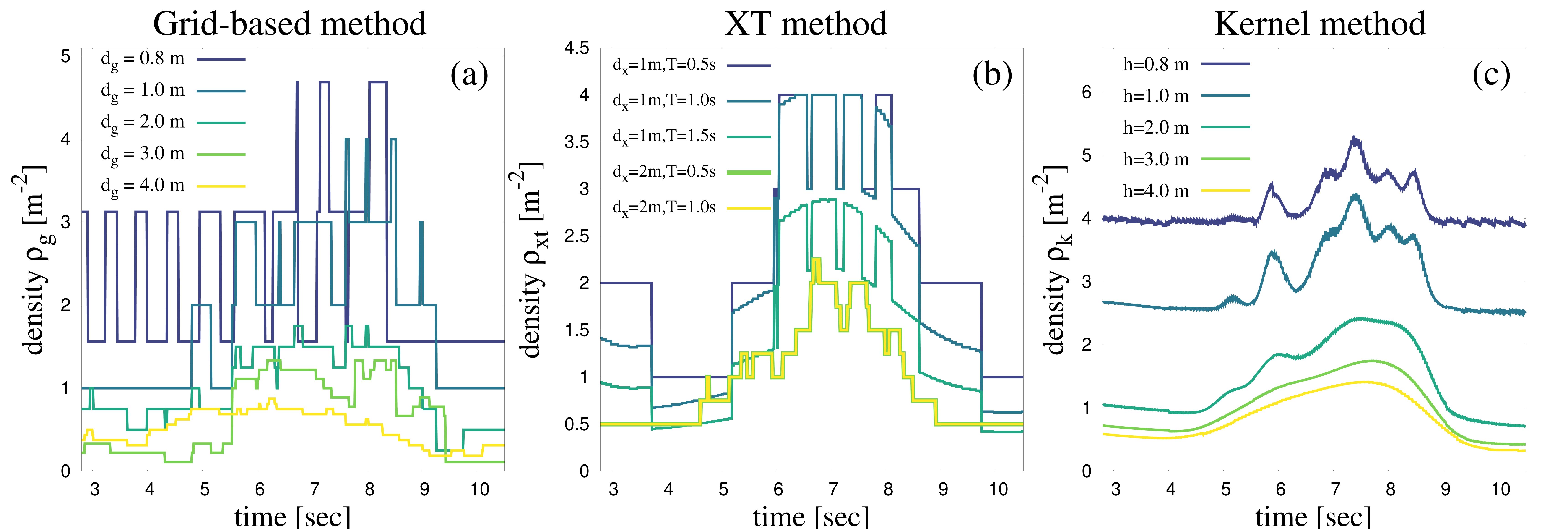}
\caption{Temporal variations of density as a function of time, along the trajectory of a given pedestrian. Density is determined by (a) the classical grid-based method, (b) the XT-method, (c) the Gaussian kernel method.  Time sequences are shown for several values of the spatial scale $d_g$, $d_x$, or $h$ and in (b), of the time window $T$.}
\label{dens_samp}
\end{figure}

In Fig. \ref{dens_samp} we show how the density along the trajectory of a given pedestrian measured by the 3 aforementioned methods varies as the spatial parameters $d_g$, $d_x$, or $h$ decrease.
In the three cases, the effect is twofold: fluctuations increase, and the average value increases towards arbitrarily high values.
While the increase of fluctuations could possibly be accounted for by more averaging, the bias introduced in the average value is more problematic and makes these methods ill-defined for a measurement along a pedestrian trajectory.

\section{Voronoi method}

It is still possible to find methods that do not have this bias along pedestrian trajectories.
These are methods which, by definition, scan the surroundings of a pedestrian on a scale such that the nearest neighbors' positions can be taken into account.

Among the various methods having this property, we shall here focus on Voronoi method - a method which is very popular due precisely to its good behavior in a large range of situations~\cite{DUIVES2015162,seyfried2010,cecile_vor,cao2017,zhang2011b}, in spite of the fact that some fluctuations are introduced by the piecewise constant nature of the density field.
We shall now describe how this method must be adapted to account for small system sizes.

The Voronoi cell of a pedestrian is the area A of the surface within which all the points are closer to that pedestrian than to any other one. The density estimate is then $\rho_v = 1/A$.

If we directly use this definition to construct the Voronoi diagram of the pedestrians at every instant of the trial, we notice that pedestrians on the edge of the groups may have a large and possibly infinite Voronoi cell.
While this indicated that these pedestrians are not constrained by neighbors at least on one side, it may be more appropriate to associate with them the density felt on the group side.

Note that if the group was confined in a space limited by walls, the latter could be used to bound the Voronoi cells. When there is no wall, a possibility is to bound Voronoi cells with an arbitrary limit. For example a restriction to 2 $m^2$ was used in~\cite{seyfried2010}. Here we prefer a method that concentrates on the density felt on the group side.

\begin{figure}[h!]
    \centering
    \includegraphics[width=\textwidth]{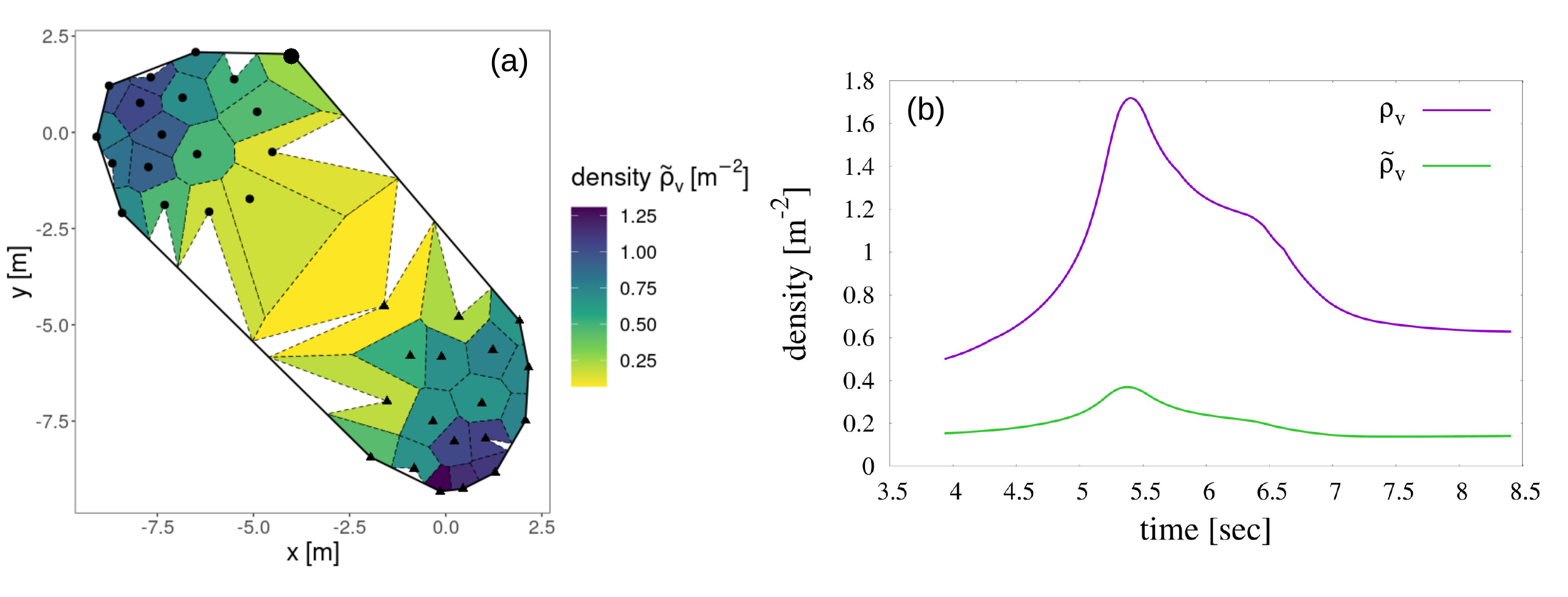}
    \caption{(a) Modified Voronoi cells for an instant from a typical trial of $\alpha = 90\degree$. The solid black line denotes the current convex hull. The dots and triangles indicate the pedestrians from two groups, which move along the $x$-axis and $y$-axis respectively. The color bar indicates the density $\tilde{\rho}_v$ experienced by each pedestrian. (b) Time sequences of the density $\rho_v$ (considering Voronoi cells clipped within the convex hull but without angular modifications) and $\tilde{\rho}_v$ (considering modified Voronoi cells clipped within the convex hull) for a typical pedestrian, denoted by a bigger black circle in (a).}
    \label{Vor_conv}
\end{figure}

To do this, a first correction already proposed in~\cite{cecile_vor} is to restrict the Voronoi cells to the angular sectors in which the Voronoi cell lies inside the convex hull encompassing the whole set of pedestrians.
Then a correction in the density accounting for the suppressed angular sectors $2\pi - \alpha$ must be performed as
\begin{equation}
    \tilde{\rho}_v = \frac{\alpha}{2\pi} \frac{1}{A}
\end{equation}
where $\alpha$ is the angular sector on which the Voronoi cell is defined.
This modification takes care of the angular adjustments for the agents located `on' the convex hull and the agents whose Voronoi cell extend beyond the convex hull.
For very small samples as the ones in our experiments, specific cases have to be taken into account.
Indeed a cell can extend on several sides of the group, requiring to suppress multiple sectors. Some examples are illustrated in Fig.~\ref{Vor_conv}(a).

 In Figure \ref{Vor_conv}(b) the variation of $\rho_v$ (considering Voronoi cells clipped within the convex hull but without angular modifications) and $\tilde{\rho}_v$ (with angular modifications) as a function of time has been plotted for a typical pedestrian. Clearly $\rho_v$ produces an overestimation, which might result in an inaccurate fundamental diagram. Hence the improvement in density estimation is brought about by the angular modification in $\tilde{\rho}_v$.


\section{Discussion and conclusion}

We have illustrated that not all methods are appropriate for density measurements along a pedestrian trajectory, and that Voronoi method, though well defined, has to be adapted in the case of small pedestrian groups.

Several other methods can also allow to measure density along a trajectory, for example using the harmonic sum of the inter-pedestrian distances from the pedestrians that are within the field of view (FOV) \cite{DUIVES2015162}, or considering the region where the convex hulls of the two crossing groups have an intersection, etc.

More generally, one must keep in mind that at small scales, a density cannot be defined in a strict way as for example for a fluid. Indeed, a proper definition of density requires that there is a scale separation between, on the one hand, the individual scale and the dimension of the area in which density is computed by a proper averaging, and, on the other hand between the latter and the scale on which density gradients occur, or on which the boundary effects become important.
This is rarely the case for pedestrians, and even less for the small samples that we consider here.

What we call density is thus rather an observable that we think relevant for the navigation of pedestrians, and that we could call {\em perceived density}, following~\cite{seyfried2010}. It certainly depends on neighboring pedestrians, but could depend more specifically on their relative positions (rear or front), on possible visual occlusions~\cite{greg2022}, on interpersonal distances~\cite{geoerg2022}, etc.
One must thus keep in mind that the choice of the {\em perceived density} plotted in the fundamental diagram actually is a hypothesis on the driving cue determining pedestrian dynamics. We shall explore various choices for this observable in our future research.

%
%
%

%

\begin{thebibliography}{10}

\bibitem{DUIVES2015162}
Duives DC, Daamen W and Hoogendoorn SP. Quantification of the level of crowdedness for pedestrian movements. Physica A \textbf{427}: 162-180 (2015).

\bibitem{seyfried2010}
Steffen B and Seyfried A. Methods for measuring pedestrian density, flow, speed and direction with minimal scatter. Physica A \textbf{389(9)}: 1902–1910 (2010)

\bibitem{cecile_vor}
Nicolas A, Kuperman M, Ibanez S, Bouzat S and Appert-Rolland C. Mechanical response of dense pedestrian crowds to the crossing of intruders. Sci. Rep. \textbf{9}: 105 (2019)

\bibitem{zhang2011b}
Zhang J, Klingsch W, Schadschneider A, Seyfried A. Transitions in pedestrian fundamental diagrams of straight corridors and t-junctions. J. Stat. Mech. p. P06004 (2011)


\bibitem{cao2017}
Cao S, Seyfried A, Zhang J, Holl S and Song W. Fundamental diagrams for multidirectional pedestrian flows. J. Stat. Mech. \textbf{2017(3)}, 033404 (mar 2017)

\bibitem{EDIE1963}
Edie  LC. Discussion of traffic stream measurements and definitions. Proceedings of the Second International Symposium on the Theory of Traffic Flow, London pp. 139–154 (1963)

\bibitem{bode_c_h2019}
Bode, NWF, Chraibi M and Holl S. The emergence of macroscopic interactions between intersecting pedestrian streams. Transp. Res. Part B-Meth. \textbf{119}: 197–210 (2019)

\bibitem{seyfried2005}
Seyfried A, Steffen B, Klingsch W and Boltes M. The fundamental diagram of pedestrian movement revisited. J. Stat. Mech. p. P10002 (2005)

\bibitem{jelic2012a}
Jelic A, Appert-Rolland C, Lemercier S and Pettre J. Properties of pedestrians walking in line – fundamental diagrams. Phys. Rev. E \textbf{85}: 036111 (2012).

\bibitem{ploscb_pratik}
Mullick P, Fontaine S, Appert-Rolland C, Olivier AH, Warren WH and Pettre J. Analysis of emergent patterns in crossing flows of pedestrians reveals an invariant of ‘stripe’ formation in human data. PLoS Comput Biol \textbf{18(6)}: e1010210 (2022).

\bibitem{pedinteract_cecile}
Appert-Rolland C, Pettre J, Olivier AH, Warren WH, Duigo-Majumdar A, Pinsard E and Nicolas A. Experimental study of collective pedestrian dynamics. Collective Dynamics \textbf{5}: 1–8 (2020).

\bibitem{guo2010}
Guo RY, Wong SC, Huang HJ, Zhang P and Lam WHK. A microscopic pedestrian-simulation model and its application to intersecting flows. Physica A \textbf{389}: 515–526 (2010)

\bibitem{zanlungo2023a}
Zanlungo F, Feliciani C, Yucel Z, Nishinari K and Kanda T. Macroscopic and microscopic dynamics of a pedestrian cross-flow: Part I, experimental analysis. Saf. Sci. \textbf{158}: 105953 (2023).

\bibitem{zanlungo2023b}
Zanlungo F, Feliciani C, Yucel Z, Nishinari K and Kanda T. Macroscopic and microscopic dynamics of a pedestrian cross-flow: Part II, modelling. Saf. Sci. \textbf{158}: 105969 (2023).

\bibitem{tordeux2015}
Tordeux A, Zhang J, Steffen B and Seyfried A. Quantitative comparison of estimations for the density within pedestrian streams. J. Stat. Mech. p. P06030 (2015).

\bibitem{greg2022}
Dachner GC, Wirth TD, Richmond E and Warren WH. The visual coupling between neighbours explains local interactions underlying human ‘flocking’. Proc. R. Soc. B \textbf{289}: 20212089 (2022)

\bibitem{geoerg2022}
Geoerg P, Schumann J, Boltes M and Kinateder M. How people with disabilities influence crowd dynamics of pedestrian movement through bottlenecks. Sci. Rep. \textbf{12}: 14273 (2022)
    
\end{thebibliography}

\end{document}